\documentclass[12pt]{iopart}

\usepackage{graphicx}
\expandafter\let\csname equation*\endcsname\relax
\expandafter\let\csname endequation*\endcsname\relax
\usepackage{amsmath}
\usepackage{bm}
\usepackage{bbm}

\usepackage{psfrag,pstricks}
\begin{document}

%......Title and other stuff
\title[Electronic properties of (111)-oriented zinc-blende quantum dots]{Electronic properties of site-controlled (111)-oriented zinc-blende InGaAs/GaAs
quantum dots calculated using a symmetry adapted $\mathbf{k}\cdot\mathbf{p}$ Hamiltonian}

\author{O. Marquardt\footnote{Present address: Paul-Drude-Institut f\"ur Festk\"orperelektronik,
Hausvogteiplatz 5 -- 7, 10117 Berlin, Germany}}
\address{Photonics Theory Group,
            Tyndall National Institute,
            Lee Maltings,
            Cork, Ireland}
\ead{marquardt@pdi-berlin.de}
\author{E.~P. O'Reilly}
\address{Photonics Theory Group,
            Tyndall National Institute,
            Lee Maltings,
            Cork, Ireland}
\address{Department of Physics,
            University College Cork,
            Cork, Ireland}
\author{S. Schulz}
\address{Photonics Theory Group,
            Tyndall National Institute,
            Lee Maltings,
            Cork, Ireland}

% 68.65.Hb: Low-dimensional, mesoscopic, nanoscale and other related systems: structure and nonelectronic properties, quantum dots
% 71.15.Dx: Methods of electronic structure calculations: Computational methodology
% 73.22.Dj: Electronic structure of nanoscale materials and related systems: Single particle states
% 78.20.Bh: Optical properties of bulk materials and thin films: Theory, models and numerical simulation
\begin{abstract}

In this work, we present and evaluate a (111)-rotated eight-band
$\mathbf{k}\cdot\mathbf{p}$ Hamiltonian for the zinc-blende crystal
lattice to investigate the electronic properties of site-controlled
InGaAs/GaAs quantum dots grown along the [111] direction. We derive
the rotated Hamiltonian including strain and piezoelectric
potentials. In combination with our previously formulated
(111)-oriented continuum elasticity model, we employ this approach
to investigate the electronic properties of a realistic
site-controlled (111)-grown InGaAs quantum dot. We combine these
studies with an evaluation of single-band effective mass and
eight-band $\mathbf{k}\cdot\mathbf{p}$ models, to investigate the
capabilities of these models for the description of electronic
properties of (111)-grown zinc-blende quantum dots. Moreover, the
influence of second-order piezoelectric contributions on the
polarisation potential in such systems is studied.
The description of the electronic structure of nanostructures grown on (111)-oriented surfaces can now be
achieved with significantly reduced computational costs in
comparison to calculations performed using the conventional
(001)-oriented models.

\end{abstract}

\date{\today}

\submitto{\JPCM}

\maketitle

\section{Introduction\label{intro}}
Quantum dots (QDs) are semiconductor structures which present quantum confinement
for charge carriers in all three spatial dimensions on the nanoscale.
They range from lithographically patterned systems of
electrons~\cite{MeHu91,WaMe92,PfGu93,EzMo97,HaSv98,Cift07} to self-assembled
nanocrystals~\cite{HiGu96,WeWa02,KiBa03,WuDo04,BrKi05,WaTh05} and QDs.
In particular, QDs from III-V semiconductor materials have attracted considerable
research interest during the past years due to their specific
electronic and optical properties that make these structures highly
promising candidates for a wide variety of novel optoelectronic
devices~\cite{Mic2003, BiGr2001,ReFo2003}. In particular, the
biexciton-exciton cascade in semiconductor QDs has been proposed as
a potential candidate for the generation of entangled photon
pairs~\cite{BeSa2000, AkLi2006}. Entangled photon pairs are a key
building block for the realisation of novel quantum-logic
applications~\cite{KnLa2001, Shie2002}. In conventional QDs grown on
the (001)-surface in zinc-blende (ZB) semiconductors, the generation
of entangled photons is difficult due to the $C_{2v}$-symmetry of
the combined system of the underlying crystal lattice and the QD
geometry~\cite{BeNa2003}. This symmetry is not high enough to allow
for a degeneracy of the bright excitonic ground
states~\cite{BeNa2003, SeSc2005}. The splitting between these states
is referred to as the fine-structure splitting
(FSS)~\cite{BeNa2003, SeSc2005}. The problem of a non-vanishing FSS
can be overcome by growing site-controlled QDs along the
[111] direction in ZB crystals~\cite{PeWa2007, ZhKa2007,JuDi13}. In this
case the symmetry of the combined system of underlying crystal
lattice and QD geometry is $C_{3v}$~\cite{SiBe2009}, which is high
enough to allow in principle for a vanishing
FSS~\cite{SiBe2009,ScWi2009}. These theoretical predictions of a
minimal FSS in ZB QDs grown along the [111] direction have been
confirmed by experiment, i.e. an extremely small FSS has been measured in
site-controlled (111)-oriented InGaAs QDs~\cite{DiMe2010,MeMa12,JuDi13,JuDi13b}.

A successful theoretical description of the electronic and optical
properties of realistic (111)-oriented site-controlled  QDs is
highly challenging~\cite{HeYo2010}, as these systems can exhibit an extremely small
aspect ratio, with base lengths as large as 50-80~nm and heights of
only 1-2~nm~\cite{MeDi2009}. This requires a
large supercell and makes atomistic calculations, e.g. employing the
empirical tight-binding method (ETBM)~\cite{SaKo2003, ScMo2009} or
the empirical pseudopotential method (EPM)~\cite{WaKi99,BeZu2005},
computationally highly expensive. Even for continuum-based
$\mathbf{k}\cdot\mathbf{p}$ methods~\cite{Bahd92,ScWi2007,FoBa2003}
the analysis of realistic (111)-oriented InGaAs QDs considered here is
computationally very demanding. This originates from the fact that
common ZB eight-band $\mathbf{k}\cdot\mathbf{p}$ models are designed
for the description of (001)-oriented systems. A major drawback of
this formulation is that the large base length of experimentally
observed (111)-grown InGaAs QDs requires a correspondingly large
supercell, whereas the small aspect ratio of these dots additionally
leads to the need for a highly accurate discretisation in all three
spatial dimensions~\cite{HeYo2010}. Therefore, when using a
(001)-oriented cell, the QD growth axis is placed along the diagonal
of the box and the size of the supercell has to be increased even
further to avoid numerical artefacts arising from the boundary
conditions. A direct strategy to significantly reduce the
computational effort to calculate the electronic properties of
(111)-oriented site-controlled InGaAs QDs is to formulate
the $\mathbf{k}\cdot\mathbf{p}$ Hamiltonian in a basis where the
[111] direction is chosen as one of the coordinate axes. This
formulation directly allows to employ a (111)-oriented supercell and
thus enables different mesh discretisations along growth- and
in-plane directions. Moreover, and in addition to computational
issues, such a formalism is also beneficial to gain deeper insight
into the key parameters that determine the electronic and optical
properties of (111)-oriented site-controlled InGaAs QDs.
As will be shown later, the fundamental weakness of the standard eight-band
$\mathbf{k}\cdot\mathbf{p}$ model, namely the inability to describe
the correct symmetry of nanostructures in the zinc-blende crystal~\cite{BeZu2005},
does not apply to (111)-oriented, $C_{3v}$-symmetric nanostructures.

For several reasons, as discussed below in detail,
$\mathbf{k}\cdot\mathbf{p}$ Hamiltonians
derived in the literature to describe (111)-oriented systems are not
directly applicable or have to be validated to be applicable to the
here studied QD systems. The effective mass approaches (EMA)
presented by Xia \emph{et al.}~\cite{Xia91} and Wei \emph{et
al.}~\cite{WeZh2010} do not take into account band mixing effects,
which have been shown to be very important for a realistic
description of the electronic structure of conventional,
(001)-oriented InGaAs QDs~\cite{WaWi2000}. Four- and six-band
Luttinger-Kohn $\mathbf{k}\cdot\mathbf{p}$ models for (111)-oriented
systems, as introduced in
Refs.~\cite{GhBa1990},~\cite{SeDo2003}
and~\cite{Kaji99}, provide a realistic description of the
valence band (VB) structure, since they take band mixing effects
into account. However, such an approach neglects the coupling
between the conduction bands (CB) and the valence bands, which
becomes important for semiconductor materials with small energy
gaps, such as InAs~\cite{WaWi2000}. Los~\emph{et al.}~\cite{LoFa96}
derived general expressions for an  eight-band
$\mathbf{k}\cdot\mathbf{p}$ Hamiltonian defined with respect to an
arbitrary orientation, taking therefore CB-VB coupling as well as
band mixing effects into account. However, an eight-band
$\mathbf{k}\cdot\mathbf{p}$ Hamiltonian for a (111)-oriented system
is not explicitly given. Moreover, most of these models are designed
to provide an accurate description of quantum wells (QWs) grown on
(111)-surfaces~\cite{Kaji99, MaSm87, IkMi92, LoFa96}, where the
effects of strain are straightforward to include in the Hamiltonian.
 However, for realistic QD geometries grown on (111)-oriented
substrates, position-dependent diagonal and off-diagonal strain
tensor components become important in particular when describing the
VBs, and therefore need to be included in the model for a consistent
description of such dots.

The aim of our present work is to provide a rotated
eight-band $\mathbf{k}\cdot\mathbf{p}$ model using a basis where the
[111] direction is chosen as one of the coordinate axes. We also
include strain and piezoelectric potentials in a similar manner as
in conventional eight-band $\mathbf{k}\cdot\mathbf{p}$
models~\cite{Bahd92,Schl2007}, designed for (001)-oriented systems.
This approach is then ideally suited to analyse the electronic
properties of realistic (111)-grown QDs with large base lengths, both in terms of computing
resource demands and also for clarity of interpretation, as we
demonstrate through the presentation of calculated electron and hole
states in a (111)-oriented InGaAs/GaAs QD structure.

The outline of this paper is as follows. In
Sec.~\ref{sec:Hamiltonian}, we derive the rotated eight-band
$\mathbf{k}\cdot\mathbf{p}$ Hamiltonian, taking spin-orbit (SO)
coupling, strain, and piezoelectric potentials into account. In
Sec.~\ref{sec:StrainPiezo_theory} we discuss the elastic energy and
the (first- and second-order) polarisation vector in a
(111)-oriented ZB system. Finally, we apply the (111)-oriented
eight-band $\mathbf{k}\cdot\mathbf{p}$ formalism, including strain
and piezoelectric fields to describe the electronic structure of a
(111)-grown In$_{0.25}$Ga$_{0.75}$As/GaAs QD of realistic dimensions
(Sec.~\ref{sec:results}). Furthermore, we study the validity of
a one-band EMA for the conduction and for the valence states.
%as well as that of a six-band $\mathbf{k}\cdot\mathbf{p}$ model for
%the valence states.
%Using the one-band EMA, we show that typical
%dots have highly asymmetric electronic properties, with few confined
%electron and many closely spaced confined hole states. We discuss
%the challenges this asymmetry presents for calculating the optical
%properties of such site-controlled QDs.
Finally, we summarise our results in Sec.~\ref{sec:summary}.

\section{Theory}
\label{sec:theory}

In this section we describe the theoretical framework employed to
analyse the electronic structure of site-controlled (111)-oriented
InGaAs/GaAs QDs. Our ansatz can be broken down into two main parts.
In the first part, Sec.~\ref{sec:Hamiltonian}, we derive expressions
for an eight-band $\mathbf{k}\cdot\mathbf{p}$ Hamiltonian adapted to
a (111)-oriented ZB system. In the second part,
Sec.~\ref{sec:StrainPiezo_theory}, we discuss the calculation of the
strain and (first- and second-order) piezoelectric potentials in
(111)-oriented ZB QDs based on our recent work~\cite{ScCa2011}.

\subsection{(111)-$\mathbf{k}\cdot\mathbf{p}$ Hamiltonian}
\label{sec:Hamiltonian}
%%%%
We derive here an eight-band $\mathbf{k}\cdot\mathbf{p}$ Hamiltonian
to describe the electronic structure of (111)-oriented ZB
structures. Our starting point is a  conventional  eight-band
$\mathbf{k}\cdot\mathbf{p}$ Hamiltonian, designed for a
(001)-oriented system, expanded using basis states with symmetry:
\begin{equation*}
\left(|S\uparrow\rangle,|X\uparrow\rangle,|Y\uparrow\rangle,|Z\uparrow\rangle,|S\downarrow\rangle,|X\downarrow\rangle,|Y\downarrow\rangle,|Z\downarrow\rangle\right)^T\,\, .
\end{equation*}
The conventional $\mathbf{k}\cdot\mathbf{p}$ Hamiltonian in this
basis is given in~\ref{ap:kphamilconv}. To obtain a
(111)-oriented eight-band Hamiltonian we proceed as described in
detail by Voon~\emph{et al.} in Ref.~\cite{VoWibook}.
Following Ref.~\cite{VoWibook}, the rotation of the
$\mathbf{k}\cdot\mathbf{p}$ Hamiltonian from the [001]- to the
[111] direction can in general be broken down into three steps. In
the first step one neglects the spin and rotates the basis functions
of the Hamiltonian. Subsequently, the un-primed wave vector
$\mathbf{k}$ and strain tensor $\epsilon$ of the (001)-oriented
system are replaced by the primed ones $\mathbf{k}'$ and $\epsilon'$
in the (111)-oriented system. In a third step, the matrix is then
re-expressed in terms of the modified basis states.

A coordinate rotation matrix $U^{c}$ going from the (001)- to the
(111)-oriented system reads~\cite{ScCa2011}:
\begin{equation}
\label{eq:RotationMatrix}
 U^{c}=
\begin{pmatrix}
 \frac{1}{\sqrt{6}} & \frac{1}{\sqrt{6}} & -\sqrt{\frac{2}{3}}\\
 -\frac{1}{\sqrt{2}} & \frac{1}{\sqrt{2}} & 0\\
 \frac{1}{\sqrt{3}} & \frac{1}{\sqrt{3}} & \frac{1}{\sqrt{3}}
\end{pmatrix}\,\, .
\end{equation}
The rotation $U^{c}$ transforms vectors $\mathbf{k}$ and tensors
$\epsilon$ from $(x, y, z)$ to $(x', y', z')$ coordinates via the
expressions~\cite{HiSi90}:
\begin{equation}
\label{eq:tranform_Basis_k}
k'_i=\sum_\alpha U^{c}_{i\alpha}k_\alpha \quad\quad , \quad\quad \epsilon'_{ij}=\sum_{\alpha,\beta}U^c_{i\alpha}U^c_{j\beta}\epsilon_{\alpha\beta}\,\, .
\end{equation}
In principle, the spatial transformation given by $U^{c}$, has to be
combined with a transformation in spin-space. However, it can be
shown that the SO coupling matrix elements are independent of the
chosen orientation of the basis states with symmetry
$(|X\rangle,|Y\rangle,|Z\rangle)^T$. This result follows from the
fact that the SO interaction is isotropic in a ZB system, which is
in contrast to $c$-plane WZ structures where the SO interaction can
be different along different directions~\cite{RoDi2001}. Therefore,
in a ZB system, the spatial rotation is already sufficient.

Following this general procedure and taking into account the
transformation rules for vectors and tensors, the eight-band
Hamiltonian $H'_\text{kp}$ of the (111)-oriented ZB structure,
expanded using basis states with symmetry
$\left(|S'\uparrow\rangle,|X'\uparrow\rangle,|Y'\uparrow\rangle,|Z'\uparrow\rangle,|S'\downarrow\rangle,|X'\downarrow\rangle,|Y'\downarrow\rangle,|Z'\downarrow\rangle\right)^T$,
can be written as:
\begin{equation}
\label{eq:full111H8}
 H'_\text{kp}=
\begin{pmatrix}
 M'(\mathbf{k}')  & \Gamma'_\text{so}\\
 -{\Gamma'}^{*}_\text{so} & {M'}^{*}(\mathbf{k}')
\end{pmatrix}
\,\, ,
\end{equation}
where $M'(\mathbf{k}')$ and $\Gamma'_\text{so}$ are both $4\times 4$
matrices. $M'(\mathbf{k}')$ is composed of matrices describing the
potential energy part $M'_\text{pe}$, the kinetic energy part
$M'_\text{ke}$, the SO interaction contribution $M'_\text{so}$ and a
strain dependent part $M'_\text{str}$:
\begin{equation}
\label{eq:full111H}
 M'(\mathbf{k}')=M'_\text{pe}+M'_\text{ke}+M'_\text{str}+M'_\text{so}\,\, .
\end{equation}
The potential energy part $M'_\text{pe}$ of $H'_\text{kp}$, which
contains terms independent of and linear in $\mathbf{k}$, is given
by:
\begin{equation}
 M'_\text{pe}=
\begin{pmatrix}
 E_\text{cb} & iPk'_x & iPk'_y & iPk'_z\\
 -iPk'_x & \tilde{E}_\text{vb} & 0 & 0\\
 -iPk'_y & 0 & \tilde{E}_\text{vb} & 0\\
 -iPk'_z & 0 & 0 & \tilde{E}_\text{vb}
\end{pmatrix}
\,\, .
\end{equation}
The CB edge is denoted by $E_\text{cb}$ while $\tilde{E}_\text{vb}$
denotes the average unstrained VB edge. $E_\text{cb}$ and
$\tilde{E}_\text{vb}$ are defined as:
\begin{eqnarray}
\label{eq:Ec}
E_\text{cb} = E_\text{vb}+V_\text{ext}+E_{g}\,\, , \,\,
\tilde{E}_\text{vb} = E_\text{vb}+V_\text{ext}-\frac{\Delta_\text{so}}{3}\,\, ,
\label{eq:Ev}
\end{eqnarray}
where $\Delta_\text{so}$ denotes the SO coupling energy, $E_{g}$ the
fundamental band gap, $E_\text{vb}$ is the averaged VB
edge on an absolute scale and $V_\text{ext}$ an optional scalar
potential describing an electric field, e.g. a piezoelectric
built-in field. The Kane parameter $P$ is defined as:
\begin{eqnarray}
\label{eq:KaneP}
P &=& \sqrt{\frac{\hbar^2}{2m_0}E_p}\,\, ,
\end{eqnarray}
where $m_0$ is the mass of an electron and
$E_p$ denotes the optical matrix element parameter.

The kinetic energy part $M'_\text{ke}$ in the (111)-oriented systems
contains the rotated VB part $h'(\mathbf{k}')$ plus the CB
contribution, which is given by ${h'}^\text{cb}=A'\mathbf{k}'^2$.
The parameter $A'$ is defined as:
\begin{eqnarray}
\label{eq:KaneA}
A'&=& \frac{\hbar^2}{2m_0}\left(\frac{1}{m_e}-\frac{E_p\left(E_g+\frac{2\Delta_\text{so}}{3}\right)}{E_g\left(E_g+\Delta_\text{so}\right)}\right)\,\, ,
\end{eqnarray}
where $m_e$ denotes the $\Gamma$-point CB effective mass. The
kinetic energy part $M'_\text{ke}$ of $H'_\text{kp}$ reads:
\begin{equation}
 M'_\text{ke}=
\begin{pmatrix}
 A'{\mathbf{k}'}^2 & 0 & 0 & 0\\
 0 & h'_{11}(\mathbf{k}') & h'_{12}(\mathbf{k}') & h'_{13}(\mathbf{k}')\\
 0 & h'_{12}(\mathbf{k}') & h'_{22}(\mathbf{k}') & h'_{23}(\mathbf{k}')\\
 0 & h'_{13}(\mathbf{k}') & h'_{23}(\mathbf{k}') & h'_{33}(\mathbf{k}')
\end{pmatrix}
\, ,
\end{equation}
with
\begin{eqnarray*}
\label{eq:vbkphamiltonian}
 h'_{11} &=& -\frac{1}{2}\left(\gamma_1+\gamma_2+3\gamma_3\right)k'^2_x-\frac{1}{2}\left(\gamma_1-\gamma_2-\gamma_3\right)k'^2_y\\
& &-\frac{1}{2}\left(\gamma_1-2\gamma_3\right)k'^2_z+\sqrt{2}\left(\gamma_2-\gamma_3\right)k'_xk'_z+\frac{P^2}{E_g}k'^2_x ,\\
 h'_{22} &=& -\frac{1}{2}\left(\gamma_1-\gamma_2-\gamma_3\right)k'^2_x-\frac{1}{2}\left(\gamma_1+\gamma_2+3\gamma_3\right)k'^2_y\\
& &-\frac{1}{2}\left(\gamma_1-2\gamma_3\right)k'^2_z-\sqrt{2}\left(\gamma_2-\gamma_3\right)k'_xk'_z+\frac{P^2}{E_g}k'^2_y ,\\
 h'_{33} &=& -\frac{1}{2}\left(\gamma_1-2\gamma_3\right)(k'^2_x+k'^2_y)-\frac{1}{2}\left(\gamma_1+4\gamma_3\right)k'^2_z\\
& &+\frac{P^2}{E_g}k'^2_z\,\, ,
\end{eqnarray*}
\begin{eqnarray*}
 h'_{12} &=& -\sqrt{2}\left(\gamma_2-\gamma_3\right)k'_yk'_z-\left(\gamma_2+2\gamma_3\right)k'_yk'_x+\frac{P^2}{E_g}k'_xk'_y,\\
 h'_{13} &=& -\frac{1}{\sqrt{2}}\left(\gamma_3-\gamma_2\right)(k'^2_x-k'^2_y)-\left(2\gamma_2+\gamma_3\right)k'_xk'_z\\
& &+\frac{P^2}{E_g}k'_xk'_z ,\\
 h'_{23} &=& -\sqrt{2}\left(\gamma_2-\gamma_3\right)k'_yk'_x-\left(2\gamma_2+\gamma_3\right)k'_yk'_z+\frac{P^2}{E_g}k'_yk'_z\,\, ,
\end{eqnarray*}
where $\gamma_i$ are the Luttinger parameters for the six-band VB
$\mathbf{k}\cdot\mathbf{p}$ Hamiltonian, defined here in units of
$\hbar^2/m_0$. The strain dependent part $M'_\text{str}$ of the eight-band
$\mathbf{k}\cdot\mathbf{p}$ Hamiltonian $H'_\text{kp}$, using
Einstein summation convention, is given by~[\footnote{
As discussed by Los~\emph{et al.},~\cite{LoFa96} in the
  presence of strain, the momentum operator $p=-i\hbar\nabla$
  transforms like \mbox{$\mathbf{p}'=(I+s)^{-1}\mathbf{p}$}, where $I$ is the
  unity matrix and $s$ the deformation tensor. Therefore, the coupling
  matrix elements $\langle S|p'_\alpha|X\rangle$, $\langle
  S|p'_\alpha|Y\rangle$ and $\langle S|p'_\alpha|Z\rangle$ with
  $\alpha=x,y,z$ are in principle also modified. This might lead to
  slightly modified effective masses for the CB and VBs~\cite{AsCa78}.
  However, Los~\emph{et al.}~\cite{LoFa96} argued
  that strain dependent coupling between CB and VBs
  is in general small. Therefore, we assume here that the Kane parameter $P$ in
  the (111)-oriented system is the same as in  the (001)-oriented
  system.}]:
\begin{equation*}
 M'_\text{str}=
\begin{pmatrix}
 a_cTr(\epsilon') & -iP\epsilon'_{1\beta}{k'}^{\beta} & -iP\epsilon'_{2\beta}{k'}^{\beta} & -iP\epsilon'_{3\beta}{k'}^{\beta}\\
 iP\epsilon'_{1\beta}{k'}^{\beta} & {h'}^\text{str}_{11} & {h'}^\text{str}_{12} & {h'}^\text{str}_{13}\\
 iP\epsilon'_{2\beta}{k'}^{\beta} & {h'}^\text{str}_{12} & {h'}^\text{str}_{22} & {h'}^\text{str}_{23}\\
 iP\epsilon'_{3\beta}{k'}^{\beta} & {h'}^\text{str}_{13} & {h'}^\text{str}_{23} & {h'}^\text{str}_{33}
\end{pmatrix}
\, .
\end{equation*}
The matrix elements ${h'}^\text{str}_{ij}$ of the strain dependent
part of the Hamiltonian can be obtained from the matrix elements
$h'_{ij}(\mathbf{k}')$  by simply using the substitutions:
\begin{eqnarray}
\label{eq:substr1}
\gamma_1 k'_ik'_j&\rightarrow& -2a_v\epsilon'_{ij}\,\, ,\\
\gamma_2 k'_ik'_j&\rightarrow& -b\epsilon'_{ij}\,\, ,\\
\gamma_3 k'_ik'_j&\rightarrow& -\frac{d}{\sqrt{3}}\epsilon'_{ij}\,\, .
\label{eq:substr2}
\end{eqnarray}
The hydrostatic VB deformation potential is denoted by
$a_v$, while $b$ and $d$ denote the uniaxial deformation potentials.

The SO related contributions $M'_\text{so}$ and $\Gamma'_\text{so}$ are given by:
%Spin orbit 1
\begin{equation}
\label{eq:SpinDiag}
 M'_\text{so}=\frac{\Delta_\text{so}}{3}
\begin{pmatrix}
 0 & 0 & 0 & 0\\
 0 & 0 & -i & 0\\
 0 & i & 0 & 0\\
 0 & 0 & 0 & 0
\end{pmatrix}
\, , \,
 \Gamma'_\text{so}=\frac{\Delta_\text{so}}{3}
\begin{pmatrix}
 0 & 0 & 0 & 0\\
 0 & 0 & 0 & 1\\
 0 & 0 & 0 & -i\\
 0 & -1 & i & 0
\end{pmatrix}
\,\, ,
\end{equation}
and are identical to the contributions in the (001)-oriented system
due to the isotropy of the SO interaction in ZB systems.

\begin{figure}[t]
\includegraphics[width=0.95\columnwidth]{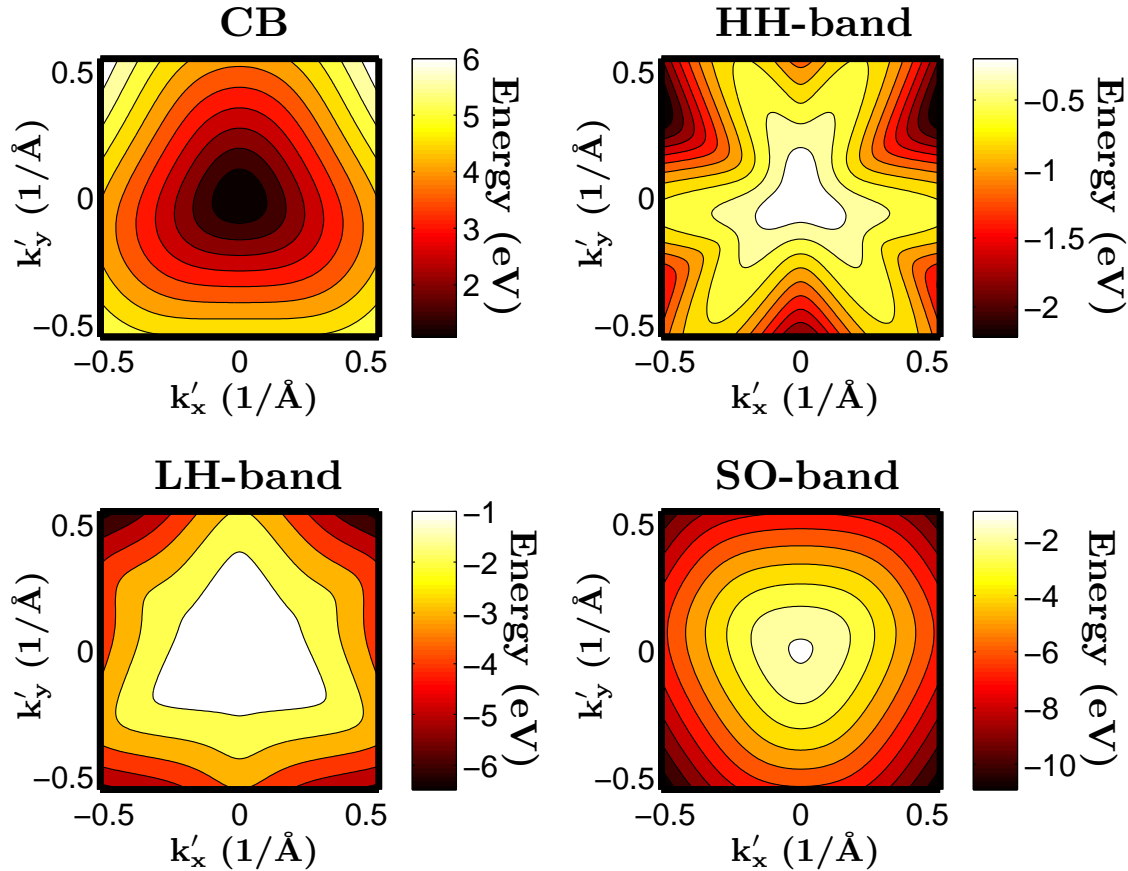}
\caption{(Colour online) Contour plots for the InAs conduction band
($\text{CB}$) and the three valence bands ($\text{VB}_1$,
$\text{VB}_2$ and $\text{VB}_3$) at $k'_z=0$ in the $k'_x$-$k'_y$-plane
obtained from the eight-band $\mathbf{k}\cdot\mathbf{p}$ Hamiltonian
$H'_\text{kp}$, equation~(\ref{eq:full111H8}).}
\label{fig:CBVBBulk_plane}
\end{figure}

\begin{figure}[t]
\includegraphics[width=0.95\columnwidth]{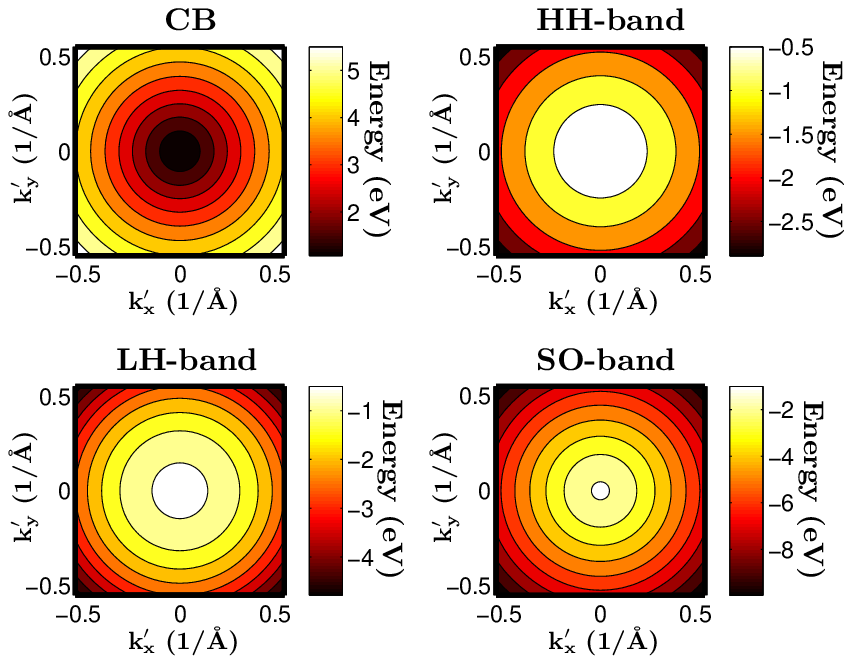}
\caption{(Colour online) Contour plots for the InAs conduction band
($\text{CB}$) and the three valence bands ($\text{VB}_1$,
$\text{VB}_2$ and $\text{VB}_3$) at $k'_z=0$ in the $k'_x$-$k'_y$-plane
obtained from the eight-band $\mathbf{k}\cdot\mathbf{p}$ Hamiltonian
$H'_\text{kp}$, equation~(\ref{eq:full111H8}), where the terms containing
\mbox{$(\gamma_2-\gamma_3)k'_ik'_j$} have been switched off.}
\label{fig:CBVBBulk_plane_iso}
\end{figure}

To illustrate the influence of the (111)-growth plane on the
conduction- and valence bands, we have calculated the CB and the VBs (HH, LH and SO-band) in the
$k'_x$-$k'_y$-plane for $k'_z=0$ . The results are shown in
figure~\ref{fig:CBVBBulk_plane} for InAs in terms of equal energy
contours. Here we find that the CB and the VBs clearly exhibit a
three-fold symmetry, as expected by considering the point group of a
ZB crystal. The tetrahedral point group $T_d$ of a ZB structure
contains clockwise and counterclockwise rotations by $2\pi/3$ around
the [111]-axis. Therefore, this real space symmetry should also be
seen in the band structure. However, the appearance of this
three-fold symmetry is also tightly linked to the difference in
Kohn-Luttinger parameters $\gamma_2$ and $\gamma_3$. In the axial
approximation, where we set the terms involving
($\gamma_2-\gamma_3$) to zero in $ M'_\text{ke}$, the three-fold
symmetry vanishes, leading to a CB and VB
dispersion which is axially symmetric about the $k'_z$ axis. This
behaviour is shown in figure~\ref{fig:CBVBBulk_plane_iso}, where we
have artificially switched off terms involving
$(\gamma_2-\gamma_3)k'_ik'_j$ in the
$\mathbf{k}\cdot\mathbf{p}$ Hamiltonian $H'_\text{kp}$,
equation~(\ref{eq:full111H8}).
%---

In order to accurately model the electronic structure of
(111)-oriented InGaAs QDs, the three-fold symmetry in CB and
VBs needs to be taken into account. It can be seen from
figure~\ref{fig:CBVBBulk_plane} that the  multiband model presented
here naturally includes the $C_{3v}$-symmetry in all directly
considered bands.
% , as required for an accurate description of
% electronic properties in these systems.

\begin{table}
\caption{Material parameters for InAs and GaAs. If not indicated otherwise, parameters are
taken from~\cite{ScWi2007}. Please note, that the
elastic constants $C_{ij}$ are calculated from the elastic constants in
(001)-oriented InAs and GaAs via the equations given in
Ref.~\cite{ScCa2011}. Interpolations of material parameters in ternary alloys follow
the equations given in~\cite{ScWi2007}.} \label{tab:materialpara} 
\begin{tabular}{|l|c|c|}
\hline
 $~$ & GaAs & InAs\\\hline
 $a$ (\AA) & $5.6503^{a}$ & $6.0553^{a}$\\
 $E_g$ (eV) & $1.518^{a}$ & $0.413^{a}$ \\
 $E_\text{vb}$ (eV) & $0.000$ & $0.173$\\
 $\Delta_\text{so}$ (eV) & $0.34^b$ & $0.38^b$ \\
 $E_p$ (eV) & 28.000 & 22.204\\
 $m_e$ ($m_0$) & 0.067 & 0.022\\
 $\gamma_1$ ($\hbar^2/m_0$) & $7.1^b$ & $19.7^b$\\
 $\gamma_2$ ($\hbar^2/m_0$) & $2.02^b$ & $8.4^b$\\
 $\gamma_3$ ($\hbar^2/m_0$) & $2.91^b$ & $9.3^b$\\
 $a_c$ (eV) & -8.013 & -5.080\\
 $a_g$ (eV) & -8.233 & -6.080\\
 $b$ (eV) & $-1.824^b$ & $-1.800^b$\\
 $d$ (eV) & $-5.062^b$ & $-3.600^b$\\
%  $C_{11}$ (GPa) & $118.8^{b}$ & $83.3^{b}$\\
%  $C_{12}$ (GPa) & $53.8^{b}$ & $45.3^{b}$ \\
%  $C_{44}$ (GPa) & $59.4^{b}$ & $39.6^{b}$ \\
 $C_{11}$ (GPa) & $140.1$ & $103.9$\\
 $C_{12}$ (GPa) & $46.7$ & $38.4$ \\
 $C_{44}$ (GPa) & $39.6$ & $25.9$ \\
 $C_{33}$ (GPa) & $147.2$ & $110.8$ \\
 $C_{13}$ (GPa) & $39.6$ & $31.6$ \\
 $C_{15}$ (GPa) & $10.0$ & $9.7$ \\
 $\epsilon_r$ & $13.18^{c}$ & $14.6^{c}$\\
 $e_{14}$ (C/m$^2$) & $-0.230^{d}$ & $-0.115^{d}$\\
 $B_{114}$ (C/m$^2$) & $-0.439^{d}$ & $-0.531^{d}$\\
 $B_{124}$ (C/m$^2$) & $-3.765^{d}$ & $-4.076^{d}$\\
 $B_{156}$ (C/m$^2$) & $-0.492^{d}$ & $-0.120^{d}$ \\
 $A_{1}$ (C/m$^2$) & $-2.656$ & $-2.894$\\
 $A_{2}$ (C/m$^2$) & $2.217$ & $2.363$\\
\hline
\multicolumn{2}{|l}{$^a$ Ref.~\cite{Bhat93}} & ~\\
\multicolumn{2}{|l}{$^b$ Ref.~\cite{Land82}} & ~\\
\multicolumn{2}{|l}{$^c$ Ref.~\cite{Adac92}} & ~\\
\multicolumn{2}{|l}{$^d$ Ref.~\cite{BeWu2006}} & ~\\
\hline
\end{tabular}
\end{table}

\subsection{Strain and Polarisation in (111)-oriented zinc-blende systems}
\label{sec:StrainPiezo_theory}

For a realistic description of the electronic structure of
site-controlled InGaAs/GaAs QDs grown on (111)-oriented substrates,
knowledge of the strain and the piezoelectric fields is required. In
principle, the stress and the electric field are coupled via the Navier
equation~\cite{VoWi2011}. However, this coupling has been shown to
be very small in InAs/GaAs systems~\cite{VoWi2011}. Therefore, we apply
here the widely used ansatz of decoupled electrostatic and elastic
equations~\cite{EdBe2007,ScWi2007,FoBa2003,Lepk2008,ScWi2009,SiBe2009,MlZu2009,ZhGo2009,McSh2010},
to study the electronic properties of semiconductor nanostructures.
In other words, our starting point for the strain field calculation
is the elastic energy $F$ of the system. Once the strain field is
known, it serves as an input for the calculation of the
piezoelectric polarisation vector and the resulting piezoelectric
built-in potential.

In this section we therefore briefly summarise the results and the
expressions we have recently obtained for the elastic energy and the
first- and second-order piezoelectric polarisation vectors in
(111)-oriented ZB structures. Combined with the rotated eight-band
$\mathbf{k}\cdot\mathbf{p}$ Hamiltonian $H'_\text{kp}$, introduced
in the previous section, this approach offers then an extremely
efficient framework to calculate the electronic structure of
(111)-oriented ZB nanostructures.

\subsubsection{Strain field calculation}

Approaches to calculate the strain fields in QD structures range
from continuum based to atomistic descriptions. Detailed discussions
of the impact of the chosen approach on the resulting strain field
have been given in a number of publications~\cite{PrKi98,StGr98}.
The choice of the strain model also depends on the choice of the
model for the electronic structure calculation. Schliwa \emph{et
al.} discussed in detail that a continuum elasticity model is the
optimal choice for an eight-band $\mathbf{k}\cdot\mathbf{p}$
approach~\cite{ScWi2007}. Furthermore, our previous analysis
showed~\cite{ScCa2011} that the use of a continuum based ansatz to
calculate the strain field in QD structures grown on a
(111)-oriented ZB substrate is already able to capture the correct
$C_{3v}$ symmetry of the system~\cite{SiBe2009}.

As described above, our starting point for the continuum based
description of the strain field in a nanostructure is the total
elastic energy $F$ of the system. To obtain the strain field in a
given nanostructure, the elastic energy $F$ of the system under
consideration is minimised with respect to the displacements
$\mathbf{u}(\mathbf{r})$. Once the displacements are known
throughout the simulation cell, the strain field can be obtained.
More details on strain field calculations in (111)-oriented ZB
systems are given in Ref.~\cite{ScCa2011}.
%--
In a second-order continuum elasticity formulation the elastic
energy $F^{(111)}_\text{ZB}$ in a uniformly strained (111)-oriented
ZB system of volume $V$ is given by~\cite{ScCa2011}:
 \begin{eqnarray}
\nonumber
  F^{(111)}_\text{ZB} &=&\frac{V}{2}\left[C_{11}(\epsilon^2_{11}+\epsilon^2_{22})+C_{33}\epsilon^2_{33}+2C_{12}\epsilon_{11}\epsilon_{22}\right. \\\nonumber
& & \left. +2C_{13}\epsilon_{33}(\epsilon_{11}+\epsilon_{22})+4C_{44}(\epsilon^2_{13}+\epsilon^2_{23})\right.\\\nonumber
& & \left. +2(C_{11}-C_{12})\epsilon^2_{12}+4C_{15}\epsilon_{13}(\epsilon_{11}-\epsilon_{22})\right.\\
& & \left. -8C_{15}\epsilon_{12}\epsilon_{23}\right]\, ,
\label{eq:F111_zb}
 \end{eqnarray}
where $\epsilon_{ij}$ denotes the different components of the strain
tensor in the (111)-oriented system while $C_{ij}$ are the
components of the stiffness tensor.
The components $C_{ij}$ have been calculated from the elastic
constants of the (001)-system according to the equations given in
Ref.~\cite{ScCa2011} and are summarised in
table~\ref{tab:materialpara}.

The expression for the elastic energy $F^{(111)}_\text{ZB}$ of a
(111)-oriented ZB structure is very similar to that for the elastic
energy of a $c$-plane WZ system~\cite{ScCa2011}. The major
difference in the above expression compared to that for the elastic
energy of a WZ system arises from the terms related to $C_{15}$.
However, it should be noted that $C_{15}$ is at least a factor of
2.5 smaller than the remaining elastic constants $C_{ij}$.
Therefore, slight modifications in the strain field of a
(111)-oriented ZB QD compared to a WZ-like system are introduced by the
$C_{15}$ related terms. A detailed discussion of strain fields in
(111)-oriented ZB structures in comparison to a WZ-like system is
also given in Ref.~\cite{ScCa2011}.

\subsubsection{First- and second-order piezoelectric polarisation}

Semiconductor materials with a lack of inversion symmetry exhibit
under applied stress an electric polarisation~\cite{Cady46}. This
strain dependent polarisation is referred to as the piezoelectric
polarisation.

In the linear regime, the piezoelectric polarisation vector is
connected to the strain state of the system via the piezoelectric
coefficient $e_{14}$. However, non-linear contributions to the
piezoelectric polarisation fields have been observed in experimental
studies on (111)-oriented InGaAs
QWs~\cite{DiMa98,ChMa2001,SaIz2002}. To take these non-linear
effects in the piezoelectric polarisation into account, different
approaches have been proposed in the
literature~\cite{BeWu2006,MiPo2006}. A detailed overview of these
different schemes is given in the recent review article by Lew Yan
Voon and Willatzen~\cite{VoWi2011}. A widely used
approach~\cite{BeZu2006,EdBe2007,ScWi2007,Lepk2008,ScWi2009,SiBe2009,MlZu2009,ZhGo2009,McSh2010},
even though it is surrounded by some
controversy~\cite{MiPo2006,VoWi2011}, was developed by
Bester~\emph{et al.}~\cite{BeWu2006}. The authors have introduced a
second-order piezoelectric tensor and calculated its coefficients.
Due to the symmetry of the system, one is left with only three
non-vanishing coefficients $B_{114}$, $B_{124}$ and $B_{156}$. The
values of these coefficients have recently been presented for the
most common III-V ZB alloys~\cite{BePr11}, including some
changes and updating of the previously reported
values~\cite{BeWu2006}. We will therefore provide a more detailed
discussion of the influence of different first- and second-order
piezoelectric constants on the polarisation potential of
(111)-oriented InGaAs QDs in Sec.~\ref{subsec:strainVp}.

In order to achieve a higher efficiency of the calculations and a
deeper insight into the key parameters that determine the
polarisation characteristics of (111)-oriented ZB QDs, we have
derived expressions for the first-
($\mathbf{P}^{(111),\text{1st}}_\text{pz}$) and
second-order
($\mathbf{P}^{(111),\text{2nd}}_\text{pz}$)
polarisation vectors in a basis where the [111] direction is chosen
as one of the coordinate axes. This basis is identical to the basis
we have used here, to obtain the eight-band
$\mathbf{k}\cdot\mathbf{p}$ Hamiltonian $H'_\text{kp}$,
equation~(\ref{eq:full111H8}), of a (111)-oriented system. The
first-order piezoelectric polarisation vector
$\mathbf{P}^{(111),\text{1st}}_\text{pz}$ is given
by~\cite{ScCa2011}:
%\begin{widetext}
\begin{equation}
\mathbf{P}^\text{(111),1st}_\text{pz} =\frac{-e_{14}}{\sqrt{3}}
\begin{pmatrix}
2\epsilon_{13}+\sqrt{2}(\epsilon_{11}-\epsilon_{22})\\
2\epsilon_{23}-2\sqrt{2}\epsilon_{12}\\
\epsilon_{11}+\epsilon_{22}-2\epsilon_{33}
\end{pmatrix}
= 2e_{14}
\begin{pmatrix}
 \sqrt{\frac{2}{3}}K_x\\
 \sqrt{2}K_y\\
 \sqrt{\frac{1}{3}}K_z
\end{pmatrix} .
% +
% \begin{pmatrix}
% e'_{11}(\epsilon'_{11}-\epsilon'_{22})\\
% 2e'_{12}\epsilon'_{12}\\
% 0
% \end{pmatrix} ,
\label{eq:FO111}
\end{equation}
%\end{widetext}
% with
% \begin{eqnarray*}
% &K_x = \frac{1}{2}\left(\epsilon_{22}-\epsilon_{11}\right)-\frac{1}{\sqrt{2}}\epsilon_{13}\,\, ,\\
% &K_y =
% \frac{1}{\sqrt{3}}\epsilon_{12}-\frac{1}{\sqrt{6}}\epsilon_{23}\;\,\,\;
% , \;\,\,\;
% % \end{eqnarray*}
% % \begin{eqnarray*}
% K_z =
% \epsilon_{33}-\frac{1}{2}\left(\epsilon_{22}+\epsilon_{11}\right)\,
% .
% \end{eqnarray*}
The strain tensor components in the (111)-ZB system
are denoted by $\epsilon_{ij}$. We introduce the coefficients $K_i$
following our previous work~\cite{ScCa2011} to simplify the
expressions for the second-order components.

In the (111)-oriented ZB system, the second-order piezoelectric
polarisation vector
$\mathbf{P}^{(111),\text{2nd}}_\text{pz}$
reads~\cite{ScCa2011}:
%\begin{widetext}
\begin{eqnarray*}
\mathbf{P}^{(111),\text{2nd}}_\text{pz}&=&2A_1\text{Tr}(\epsilon)
\begin{pmatrix}
 \sqrt{\frac{2}{3}}K_x\\
 \sqrt{2}K_y\\
 \sqrt{\frac{1}{3}}K_z
\end{pmatrix}
 +2A_2
\begin{pmatrix}
 \sqrt{\frac{2}{3}}\left[C_1\left(K_z-K_x\right)+C_2K_y\right]\\
 \sqrt{2}\left[C_2C_3+C_1K_y\right]\\
 \sqrt{\frac{1}{3}}\left[2C_1K_x+2C_2K_y\right]
\end{pmatrix}
\end{eqnarray*}
\begin{eqnarray}
~&+&4B_{156}
\begin{pmatrix}
\sqrt{\frac{2}{3}}\left[K_y^2-C_3K_x\right]\\
\sqrt{2}\left[-K_yC_4\right]\\
\sqrt{\frac{1}{3}}\left[C_3\left(K_z-K_x\right)-K_y^2\right]
\end{pmatrix}\, ,
\label{eq:111SOPolaVec}
\end{eqnarray}
%\end{widetext}
%-------------
with
\begin{eqnarray*}
C_1 &=& \frac{1}{4}(\epsilon_{22}-\epsilon_{11})+\frac{1}{\sqrt{2}}\epsilon_{13}\,\, ,\\%-\sqrt{\frac{3}{2}}\epsilon'_{23}-\frac{\sqrt{3}}{2}\epsilon'_{12}\right]\\
%--
% % \end{eqnarray*}
% % \begin{eqnarray*}
% \end{eqnarray*}
% \begin{eqnarray*}
C_2 &=& \sqrt{\frac{3}{2}}\epsilon_{23}+\frac{\sqrt{3}}{2}\epsilon_{12}\,\, ,\\%\frac{1}{3}\left(\epsilon'_{33}-\epsilon'_{11}\right)-\frac{1}{3\sqrt{2}}\epsilon'_{13}\\
C_3 &=&
\frac{1}{3}(\epsilon_{33}-\epsilon_{11})-\frac{1}{3\sqrt{2}}\epsilon_{13}\,\,
,
\end{eqnarray*}
and
\begin{eqnarray*}
C_4 &=&
\frac{1}{6}\epsilon_{11}-\frac{1}{2}\epsilon_{22}+\frac{1}{3}\epsilon_{33}+\frac{\sqrt{2}}{3}\epsilon_{13}\,\,
.
\end{eqnarray*}
The coefficients $A_1$ and $A_2$ are related to the piezoelectric
coefficients $B_{114}$ and $B_{124}$  by
\begin{equation}
\label{eq:A1A2}
 A_1 = \frac{1}{3}\left(B_{114}+2B_{124}\right)~\mathrm{and}~A_2 =
 \frac{2}{3}\left(B_{114}-B_{124}\right) ,
\end{equation}
with the coefficients $A_1$ and $A_2$ describing the second order
piezoelectric response associated respectively with hydrostatic and
biaxial strain. We note that the components of the $A_1$
second-order piezoelectric vector (hydrostatic strain term) in
equation~(16) are directly proportional to those for the first order term
in equation~(15).
% However, we present the first order vector in terms of
% the strain tensor components $\epsilon_{ij}$ and the second order
% expression in terms of the vector components $K_i$ for consistency
% with the notation used in Ref.~\cite{ScCa2011}.

Once $\mathbf{P}^{(111),\text{1st}}_\text{pz}$ and
$\mathbf{P}^{(111),\text{2nd}}_\text{pz}$ are known
for a given structure, the corresponding piezoelectric potentials
can be calculated by solving the Maxwell equation $\nabla\cdot
\mathbf{D}=0$. More details are given in Ref.~\cite{ScCa2011}.

\section{Results}
\label{sec:results}

In this section we present our results on the electronic structure
of site-controlled (111)-oriented InGaAs QDs. In a first step we
review experimental data on the structural properties of these
systems. We briefly discuss the strain and electrostatic built-in
fields in these systems and point out some similarities and
differences to conventional (001)-oriented InGaAs QDs. A more
detailed discussion is given in Ref.~\cite{ScCa2011}. Finally,
in Sec.~\ref{subsec:single-particle}, we focus on the
single-particle electron and hole states in a realistic
site-controlled (111)-oriented InGaAs QD.

\subsection{Model geometry}
\label{subsec:QDgeometry}

\begin{figure}[t]
\includegraphics[width=0.5\columnwidth]{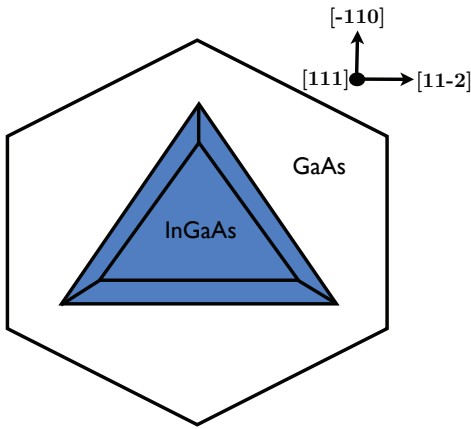}
\caption{(Colour online) Schematic view of a (111)-oriented InGaAs QD in a hexagonal,
(111)-oriented simulation cell.} \label{fig:schematic}
\end{figure}

Experimental data on site-controlled (111)-oriented InGaAs QDs
indicate a triangular-shaped QD geometry~\cite{PeWa2004}. As
discussed above, these triangular-shaped QDs exhibit a very small
aspect ratio with base lengths of order 50-80 nm and heights of only
1-2 nm~\cite{HeYo2010}. The reported indium concentration in the
site-controlled (111)-oriented InGaAs QDs considered here ranges from
15\% to 45\%~\cite{HeYo2010}. Therefore, in accordance with these
experimental findings, we choose a triangular
In$_{0.25}$Ga$_{0.75}$As QD grown on the (111)-surface with a
triangle side length of 80~nm and a height of 2~nm. The QD is
embedded in a GaAs matrix. A schematic illustration of such
a QD is shown in Fig.~\ref{fig:schematic}.
%--------
%--------

Making use of the advantages of the
(111)-rotated $\mathbf{k}\cdot\mathbf{p}$ model introduced above, we
have discretised our simulation cell in steps of 4~nm in-plane and
0.5~nm along the growth direction (the [111] direction), to achieve a reasonable
computational effort for our calculations. Due to the plane-wave
framework of the S/Phi/nX~\cite{MaBo2010,BoFr2011} software package
used for our calculations, periodic boundary conditions are assumed.
All material parameters employed in our calculations are listed in
table~\ref{tab:materialpara}.
To increase the efficiency of these calculations in a (111)-oriented cell,
we have moreover made use of a hexagonal shaped supercell, where the symmetry of the
simulation cell does not interfere with the $C_{3v}$ symmetry of the combined system
of the underlying lattice and QD geometry.
%  be further improved by using
%triangular- or hexagonal shaped supercells. Such a supercell would
%not interfere with the $C_{3v}$ symmetry of the combined system of
%underlying lattice and QD geometry. However, our main interest here
%is to set up a coherent theoretical framework in which all
%calculations can be performed in a (111)-oriented cell and to gain
%some first insights into the electronic structure of realistic
%site-controlled (111)-oriented InGaAs QDs. Therefore, we do not
%perform calculations for different In compositions or QD geometries
%and use a cubic supercell with periodic boundary conditions.
%To minimise numerical artefacts arising from the chosen boundary
%conditions, especially accounting for the long-range effects of
%strain and polarisation potential, a cell size of
%500$\times$500$\times$30~nm$^3$ has been chosen\textcolor{red}{, which corresponds to
%approx. 330~Mio. atoms, and which is beyond the capabilities of modern atomistic
%simulation software such as tight-binding or empirical pseudopotential models.}
%---
Of course the question of the electronic structure
of site-controlled (111)-oriented InGaAs QD can, in principle, also
be addressed by means of EPM or ETBM approaches. However, as in the
case of conventional $\mathbf{k}\cdot\mathbf{p}$-Hamiltonians,
simulation packages such as Nemo3D~\cite{KlOy2002} (ETBM) are primarily
designed for (001)-oriented ZB structures. Therefore, the
underlying atomic grid has to be adjusted for the system under
consideration. Furthermore, as the experimentally
observed QDs exhibit large base lengths of up to 80~nm, and because the
dots are considered as isolated systems, it is required to provide a
sufficiently large unit cell around the QD, in order to avoid
artefacts arising from the long-range
piezoelectric potential or strain. Thus, supercells employed can
easily exceed dimensions of 200~nm along the in-plane directions and
50~nm along the [111] growth direction, since
hexagonal or triangular-shaped cells are not the standard
implementation in most simulation packages. Therefore, supercells
dimensions of at least $200\times200\times50 \text{nm}^3$ are required
for the QDs considered here, which corresponds
to almost 90~million atoms, making more sophisticated atomistic
calculations extremely cumbersome and time-consuming, with
considerable requirements for computational resources.
The benefit of the here presented
$\mathbf{k}\cdot\mathbf{p}$-Hamiltonian for (111)-oriented ZB
systems in conjunction with the S/Phi/nX software library
is that a ready-to-use solution to the problem of the electronic structure of
(111)-oriented InGaAs QDs is provided. The $\mathbf{k}\cdot\mathbf{p}$ module of the S/Phi/nX package
has been developed to accept any arbitrary
$N$-band Hamiltonian as an input file, where the Hamiltonian is set
up in an input file in a human readable
meta-language~\cite{MaSc2012}. Therefore, no additional coding is
required once the desired $\mathbf{k}\cdot\mathbf{p}$ Hamiltonian has been prepared.

\subsection{Strain and polarisation potential}
\label{subsec:strainVp}

As known from (001)-oriented InGaAs QDs, strain and built-in fields
significantly modify the electronic and optical properties of
epitaxially grown QD systems. Recently, we have discussed in detail
the strain and built-in fields in (111)-oriented InGaAs
QDs~\cite{ScCa2011}. Our results have shown that the strain, first-
and second-order built-in fields exhibit a three-fold symmetry
($C_{3v}$) even if the QD geometry possesses a higher symmetry, e.g.
$C_{\infty}$ symmetry. This symmetry is further emphasised by
the triangular shape of realistic site-controlled (111)-oriented
InGaAs QDs. Moreover, and in contrast to (001)-oriented InGaAs QDs,
one finds a potential drop along the growth direction of the
nanostructure. This behaviour is similar to a nitride-based WZ
structure~\cite{WiSc2009}. However, the potential drop in a realistic
nitride WZ nanostructure is much larger due to the much larger
piezoelectric coefficients and the spontaneous polarisation which is
missing in ZB systems~\cite{ScCa2011}. This potential drop also affects the
single-particle states of (111)-oriented InGaAs QDs. Depending on
the dot height and In concentration, this effect might lead to a
spatial separation of electron and hole wave functions, diminishing
therefore the oscillator strength of interband transitions.
Furthermore, as shown by Schliwa ~\emph{et al.}~\cite{ScWi2007} for
(001)- and (111)-oriented InGaAs QDs, the balance between first- and
second-order piezoelectric contributions is very sensitive to the QD
shape, size and composition. Therefore, the detailed theoretical
analysis of the electronic and optical properties of high-quality
site-controlled (111)-oriented InGaAs QDs in combination with
experimental studies provides a very promising route to gain deeper
insight into non-linear piezoelectric effects in ZB
semiconductor materials.

Following our discussion in Sec.~\ref{sec:StrainPiezo_theory}, we
have calculated the polarisation potential in our model QD using
only first-order and first- plus second-order contributions to
evaluate the influence of second-order piezoelectric effects on the
polarisation potential of the system. Modified values have recently
been presented~\cite{BePr11} for the  first- and second-order
piezoelectric constants for InAs and GaAs originally reported in
Ref.~\cite{BeWu2006}. In particular, the $B_{156}$ parameter
differs from the previously reported value. This parameter changes
its sign for InAs, compared to the previous value. The sign remains
the same in GaAs, but the absolute value differs from the previous
parameter set by about $40\%$. However, as we have demonstrated
recently, for the total built-in potential in (111)-oriented
InAs/GaAs QDs, contributions arising from $B_{156}$ terms are of
secondary importance~\cite{ScCa2011}. Moreover, the changes in the
dominant parameters $A_1$ and $A_2$, equation~(\ref{eq:A1A2}), are $\leq
2$\% [cf. table~\ref{tab:diffVp}]. Therefore, only slight changes in
the total built-in potential are expected in (111)-oriented
InGaAs/GaAs QDs when using the two different parameter sets. However,
for the sake of completeness, our calculations employing both first-
and second-order contributions were carried out with the two
different parameter sets. In doing so, we were able to identify the
quantitative influence of the different parameter sets on the
polarisation potential. Overall, we find second-order piezoelectric
effects to have a non-negligible influence on the polarisation
potentials. However, as expected from the discussion above, only
minor modifications on the polarisation potential arise from the
choice of different parameter sets for the second-order
piezoelectric contributions. In particular, we have not seen any
qualitative change of the polarisation potential in our QD system
and quantitative changes are small as the minima and maxima of the
resulting polarisation potential, shown in table~\ref{tab:diffVp},
indicate. Based on this evaluation, we have performed our
calculations using the first- and second-order piezoelectric
constants from Ref.~\cite{BeWu2006}.
\begin{table}
\caption{First- and second-order piezoelectric constants $e_{14}$,
$B_{114}$, $B_{124}$, $B_{156}$ and the resulting $A_1$ and $A_2$
from different references as well as the corresponding minima and maxima of
the polarisation potential $V_p$.}
 \label{tab:diffVp}
 \begin{tabular}{|l|c c|c c|c c|}
 \hline
 ~ & \multicolumn{2}{c|}{Ref.~\cite{Adac92}} &  \multicolumn{2}{c|}{Ref.~\cite{BeWu2006}} &  \multicolumn{2}{c|}{Ref.~\cite{BePr11}}\\
 Parameter & InAs   & GaAs  & InAs  & GaAs  & InAs  & GaAs\\
 \hline
$e_{14}$~[$\text{C/m}^2$]  & -0.045 & -0.160 & -0.115 & -0.230 & -0.115 & -0.238\\
\hline
$B_{114}$~[$\text{C/m}^2$]     & \multicolumn{2}{c|}{n.a.} & -0.531    & -0.439    & -0.6  & -0.4\\
$B_{124}$~[$\text{C/m}^2$]     & \multicolumn{2}{c|}{n.a.} & -4.076    & -3.765    & -4.1  & -3.8\\
$B_{156}$~[$\text{C/m}^2$]     & \multicolumn{2}{c|}{n.a.} & -0.120    & -0.492    & 0.2   & -0.7\\
\hline
$A_1$~[$\text{C/m}^2$] & \multicolumn{2}{c|}{n.a.} & -2.894    & -2.656    & -2.933    & -2.667\\
$A_2$~[$\text{C/m}^2$] & \multicolumn{2}{c|}{n.a.} & 2.363 & 2.217 & 2.333 & 2.267\\
\hline
min($V_p$)~[mV]   & \multicolumn{2}{c|}{-35.9}    & \multicolumn{2}{c|}{-38.1}    & \multicolumn{2}{c|}{-39.8}\\
max($V_p$)~[mV]   & \multicolumn{2}{c|}{28.3} & \multicolumn{2}{c|}{30.4} & \multicolumn{2}{c|}{31.8}\\
\hline
\end{tabular}
\end{table}

\begin{figure}[htp]
\begin{tabular}{|c|c|c|c|}
\hline
\hline
$|\Psi^0_e|^2$ & $|\Psi^1_e|^2$ & $|\Psi^2_e|^2$ & $|\Psi^3_e|^2$ \\
\hline
\includegraphics[width=.15\textwidth]{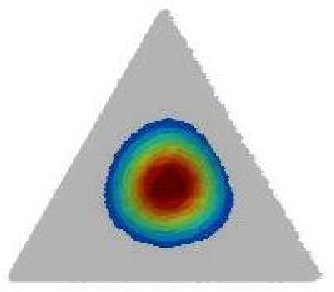} &
\includegraphics[width=.15\textwidth]{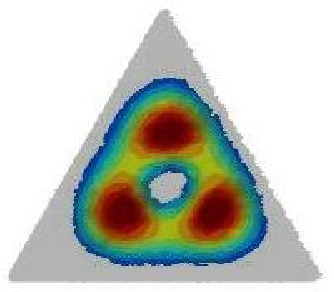} &
\includegraphics[width=.15\textwidth]{el12.eps} &
\includegraphics[width=.15\textwidth]{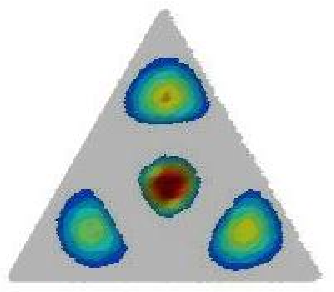}\\
\hline
$E_e^0$ = 1491.4 meV & $E_e^1$ = 1497.4 meV & $E_e^2$ = 1497.4 meV & $E_e^3$ = 1503.6 meV\\
\hline
$|X'\pm\rangle$: 0.0012 & $|X'\pm\rangle$: 0.0027 & $|X'\pm\rangle$: 0.0026 & $|X'\pm\rangle$: 0.0039 \\
$|Y'\pm\rangle$: 0.0012 & $|Y'\pm\rangle$: 0.0026 & $|Y'\pm\rangle$: 0.0027 & $|Y'\pm\rangle$: 0.0040 \\
$|Z'\pm\rangle$: 0.0107 & $|Z'\pm\rangle$: 0.0110 & $|Z'\pm\rangle$: 0.0109 & $|Z'\pm\rangle$: 0.0105 \\
$|S'\pm\rangle$: 0.9870 & $|S'\pm\rangle$: 0.9830 & $|S'\pm\rangle$: 0.9840 & $|S'\pm\rangle$: 0.9818 \\
\hline
\hline
$|\Psi^0_h|^2$ & $|\Psi^1_h|^2$ & $|\Psi^2_h|^2$ & $|\Psi^3_h|^2$ \\
\hline
\includegraphics[width=.15\textwidth]{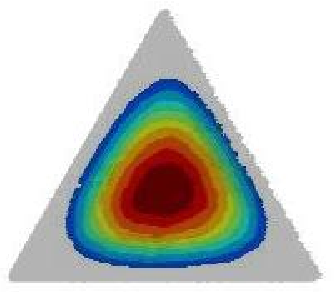} &
\includegraphics[width=.15\textwidth]{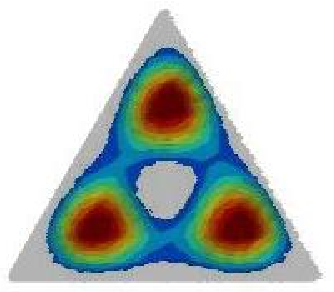} &
\includegraphics[width=.15\textwidth]{ho12.eps} &
\includegraphics[width=.15\textwidth]{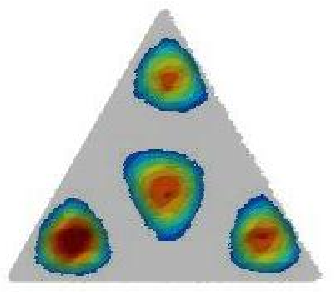}\\
\hline
$E_h^0$ = 64.9 meV & $E_h^1$ = 63.5 meV & $E_h^2$ = 63.4 meV & $E_h^3$ = 61.2 meV\\
\hline
$|X'\pm\rangle$: 0.4989 & $|X'\pm\rangle$: 0.4969 & $|X'\pm\rangle$: 0.4950 & $|X'\pm\rangle$: 0.4933 \\
$|Y'\pm\rangle$: 0.4976 & $|Y'\pm\rangle$: 0.4940 & $|Y'\pm\rangle$: 0.4970 & $|Y'\pm\rangle$: 0.4927 \\
$|Z'\pm\rangle$: 0.0035 & $|Z'\pm\rangle$: 0.0081 & $|Z'\pm\rangle$: 0.0069 & $|Z'\pm\rangle$: 0.0121 \\
$|S'\pm\rangle$: 0.0001 & $|S'\pm\rangle$: 0.0011 & $|S'\pm\rangle$: 0.0011 & $|S'\pm\rangle$: 0.0018 \\
\hline 
\hline
\end{tabular}
\caption{(Colour online) Top view of the four electron (top) and hole (bottom) states closest to the band gap in a 2.0~nm thick In$_{0.25}$Ga$_{0.75}$As QD
with a base length of 80~nm,
calculated using a symmetry adapted eight-band $\mathbf{k}\cdot\mathbf{p}$ Hamiltonian. The dot is marked in gray. All energies are given with
respect to the bulk GaAs valence band edge.
Additionally, the orbital contributions averaged over spin up and down ($|\pm\rangle$) are indicated.}\label{fig:el111}
%\caption{(Colour online) Charge densities of the electron
%($|\Psi^0_e|^2$) and hole ground state ($|\Psi^0_h|^2$) along with
%the first three excited hole states ($|\Psi^{1,2,3}_h|^2$)
%relative to the unstrained GaAs VB edge.
%The QD is
%shown from above looking along the [111] direction. Bright (blue)
%and dark (purple) isosurfaces represent 30$\%$ and 60$\%$ of the
%maximum charge density, respectively.} \label{tab:states}
\end{figure}

\subsection{Single-particle states}
\label{subsec:single-particle}

The formalism derived in Sec.~\ref{sec:Hamiltonian} has been
employed and evaluated for the description of the electronic
properties of a (111)-oriented, site-controlled
In$_{0.25}$Ga$_{0.75}$As/GaAs QD of triangle side length 80~nm and
of height 2~nm.
Figure~\ref{fig:el111} shows the eigenenergies, charge densities
and orbital contributions of the first four localised electron and hole states.
Please note, each state is twofold Kramer's degenerate.
The electron states are dominated by the $|S'\uparrow\rangle$ and $|S'\downarrow\rangle$ states
with only small contributions from the $|X'\uparrow\downarrow\rangle$, $|Y'\uparrow\downarrow\rangle$,
and $|Z'\uparrow\downarrow\rangle$ bands.
All four hole states depicted in Fig.~\ref{fig:el111} are predominantly HHs with large
$|X'\uparrow\rangle$, $|Y'\uparrow\rangle$, $|X'\downarrow\rangle$
and $|Y'\downarrow\rangle$ contributions to the eigenstates, and only a small contribution
from $|Z'\uparrow\rangle$, $|Z'\downarrow\rangle$,
$|S'\uparrow\rangle$ and $|S'\downarrow\rangle$ components.
%For the chosen QD geometry and material composition one finds only one bound electron state.
%The difference in the number of bound states arises from the difference
%in the effective masses of electrons and holes along the
%[111] direction. Since the holes have a higher effective mass
%($m_{hh}^{[111]}\approx 0.825\,m_0$) than the electrons
%($m_{e}\approx 0.051\,m_0$) along this direction, the electron
%single-particle states and energies are more sensitive to the height
%and width of the confinement potential in the [111] direction,
%in particular as the CB edge inside the
%QD is shifted to higher energies by the strong deformation potential
%$a_c$ in InAs and GaAs (cf. table~\ref{tab:materialpara}).
%--------
\begin{figure}[t]
\includegraphics[width=0.85\columnwidth]{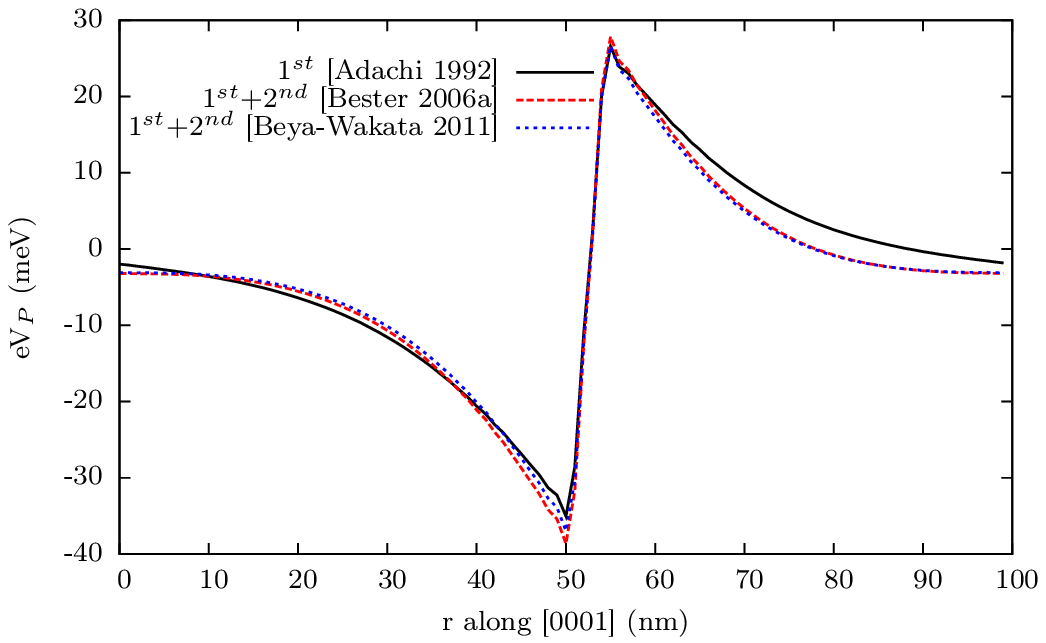}
\caption{(Colour online) Polarisation potential energy shown as a
line scan through the QD centre for the different first- and second-order piezoelectric
constants shown in Table~\ref{tab:diffVp}.} \label{fig:vpol}
\end{figure}
%--------

Additionally, a spatial separation of electron and hole
charge densities is observed, resulting from the polarisation potential
shown in figure~\ref{fig:vpol}, where a potential drop is visible, similar
to those in WZ QDs along the growth direction. 
%The weak electron localisation as well as the spatial separation of the
%charge carriers  can be seen more clearly in
%figure~\ref{fig:states1d}, where the charge density of the electron
%and hole ground state obtained from the eight-band
%$\mathbf{k}\cdot\mathbf{p}$ model is shown for a line-scan along the
%[111] direction through the QD centre. It is clearly visible, that
%the localisation of the electron state inside the QD is much weaker
%than that of the hole state.
%We note that for an ideal, isolated triangular (111)-oriented QD,
%symmetry requires that the states depicted in figure~\ref{fig:states}
%should have equal $|X'\rangle$ and $|Y'\rangle$ character; the small differences
%in the presented composition of the wave functions here reflect the
%influences of the cubic supercell geometry chosen.
As a result of the flat shape of the QD
with a height of only 2~nm and the correspondingly weak charge
carrier localisation, the electron ground state binding energy
is quite close to the GaAs CB,
resulting in a calculated ground state single-particle transmission energy of approx. 1428.3~meV.
This value is smaller than the one reported by Juska et al.~\cite{JuDi2011} of 1462.5~meV. However, our model QD
has a larger thickness than the one reported in this reference, which leads to lower quantisation energies in
our system.
%As excitonic effects have not been considered here and possible effects
%such as alloy fluctuations might induce further deviations, this
%result is in good agreement with experimental observations on such
%systems, where emission energies of \textbf{TODO: Stefan: bitte den aktuellen Wert von Gediminas eintragen! ?1441?} meV were
%observed~\cite{MeDi2009}.
Previous measurements on site-controlled InGaAs/GaAs QDs reported smaller emission energies of around
1304~meV~\cite{GaFe08}, resulting from different dimensions (base
lengths of 20~nm and heights of 5~nm), and material
composition. When looking at the charge densities in more
detail (Figure~\ref{fig:el111}), one finds that the electron and hole
ground state wave functions can be classified according to their
nodal structure as being $s$-like.
%--------
%\begin{figure}[t]
%\includegraphics[width=0.85\columnwidth]{./Fig5.eps}
%\caption{(Colour online) Electron (solid red) and hole (dash-dotted
%blue) ground state charge densities plotted along the
%[111] direction throughout the QD centre. The QDs interfaces are
%marked with thin black dashed lines.} \label{fig:states1d}
%\end{figure}
%--------
The following excited electron states exhibit p-like nodal structures.
First and second excited states are moreover
degenerate as a consequence of the system's $C_{3v}$ symmetry.
However, the classification of the first three excited hole states
is more complicated due to the underlying $C_{3v}$ symmetry of the
system. The first two excited hole states, $\Psi_h^1$ and $\Psi_h^2$, neglecting the Kramer's
degeneracy of these states, are almost degenerate. From the 
$C_{3v}$ symmetry of the system one might have expected an exact
degeneracy. Since our calculations take SO coupling effects
into account, one has to deal with the \emph{double} group
$\bar{C}_{3v}$ which allows for twofold degenerate states
only~\cite{ScSc2008}. If we artificially switch off the SO coupling
in our calculations and neglect strain and polarisation potentials,
the states $\Psi^1_h$ and $\Psi^2_h$ are indeed degenerate. Strain
and polarisation potentials calculated from a continuum-elasticity
model as described in Ref.~\cite{ScCa2011} do not reduce the symmetry of our model QD and thus do not
induce a splitting of degenerate states.
%However, the periodic
%boundary conditions of our plane-wave based formalism in a cubic
%supercell lead to a slight asymmetry, that breaks existing
%degeneracies for this model system. These effects can be fully
%removed by employing a supercell that matches the QD and crystal
%symmetry~\cite{MoBa10}.

% --- new part on effective mass, 6-band etc.

The computational effort of simulating  electronic properties in a
(111)-oriented QD with such an extremely small aspect ratio can be
further reduced strongly by employing a one-band EMA instead of a
multiband $\mathbf{k}\cdot\mathbf{p}$ approach. Moreover, the large
base length and the small height of the QD suggest an almost QW-like
behaviour, such that an EMA might in fact be suited to provide an
accurate description of the confined states in such a system. To
evaluate the applicability of simplified EMA models, we compare now the results
of the single-band effective mass model with those from an eight-band
$\mathbf{k}\cdot\mathbf{p}$ approach used to describe the electronic
properties of our model system.
% The two-plus-six-band model for the
%description of the hole states can be obtained from the eight-band
%model by setting the Kane parameter $P$ to zero.
In the single-band effective mass case, we consider an isotropic electron mass
$m_e(\mathbf{r})$:
\begin{equation}
\hat{H}_e=\frac{\hbar^2}{2
m_0}\frac{1}{m_e(\mathbf{r})}\mathbf{k}'^2 +
E_\mathrm{cb}(\mathbf{r}) + a_c(\mathbf{r})\cdot
Tr(\epsilon(\mathbf{r})) + V_\text{ext}(\mathbf{r}).
\end{equation}
The hole effective mass is different along the [111]-growth direction $z'$ and in the
in-plane directions $x'$ and $y'$:
\begin{eqnarray}
\nonumber \hat{H}_h &=&
- \frac{1}{2}\cdot\left[(\gamma_1(\mathbf{r}) -
2\gamma_3(\mathbf{r}))\cdot k_z'^2
\right.\\
\nonumber & & \left.
+ (\gamma_1(\mathbf{r}) + \gamma_3(\mathbf{r}))\cdot(k_x'^2 + k_y'^2)\right]+ \tilde{E}_\mathrm{vb}(\mathbf{r})\\
\nonumber & & + a_v(\mathbf{r})\cdot Tr(\epsilon) +
V_\text{ext}(\mathbf{r})\\
& & - \frac{d(\mathbf{r})}{\sqrt{3}}\cdot
\left[\epsilon_{33}(\mathbf{r})-\frac{1}{2}(\epsilon_{11}(\mathbf{r})+\epsilon_{22}(\mathbf{r}))\right]\,.
\end{eqnarray}

%When using simplified two-plus-six band models or EMA, 
%the nodal structure of the electron and hole states remains similar to that
%calculated using the eight-band model. However, the eigenenergies of
%these states are affected by the choice of simplified models.
Table~\ref{tab:multibandEnergies} summarises the electron and hole
eigenenergies obtained from the two models with respect to the bulk GaAs VB edge.
\begin{table}
\caption{Electron and hole eigenenergies for eight-band
$\mathbf{k}\cdot\mathbf{p}$ model and an EMA in meV relative to the
GaAs VB edge. Note that within the eight-band
model each state is twofold degenerate due to time reversal symmetry.}
 \label{tab:multibandEnergies}
 \begin{tabular}{|l|c c c c|c c c c|}
 \hline
 model & $E(\Psi_e^0)$ & $E(\Psi_e^1)$ & $E(\Psi_e^2)$ & $E(\Psi_e^3)$ & $E(\Psi_h^0)$ & $E(\Psi_h^1)$ & $E(\Psi_h^2)$ & $E(\Psi_h^3)$\\
 \hline
% 8-band &   1495.9  & 1501.9   & 1501.9   & 1508.1 & 63.7 & 62.4 & 62.2 & 60.1\\ % old vals
% EMA    &   1496.9  & 1503.2   & 1503.2   & 1509.5 & 62.4 & 59.5 & 59.5 & 55.3\\ % old vals
 8-band &   1491.4  & 1497.4   & 1497.4   & 1503.6 & 64.9 & 63.5 & 63.4 & 61.2\\ % new vals, 25% In
 EMA    &   1492.4  & 1498.7   & 1498.7   & 1505.0 & 63.6 & 60.6 & 60.6 & 56.4\\ % new vals, 25% In
 \hline
\end{tabular}
\end{table}
%It can be seen, that the two-plus-six-band model is in excellent
%agreement with the eight-band approach for the system under
%consideration. Differences in the hole state eigenenergies are in
%the order of less than 1~meV. While this good agreement holds for
%our model system, stronger CB-VB coupling can be expected for
%systems with higher In-content, different QD geometries, or other ZB
%semiconductor materials, and is thus not generally conclusive.
%Furthermore, the computational effort is not significantly higher in
%an eight-band model compared to a six-band one.
It can be seen that the EMA-based results are in broad agreement with but show
some deviations from the eight-band model. In
addition, the EMA neglects the SO coupling, such that the first and
second excited hole state are expected and found to be energetically
degenerate, which can represent a significant difference to results obtained
from an eight-band model, depending on QD geometry and composition.
%is a significant difference to the results
%obtained from the eight-band model.
%This is basically what we find;
%however, a small splitting remains between these two hole states in
%the EMA because of  minor asymmetries in the polarisation potential
%and strain field, due to the periodic boundary conditions assumed.

While the charge densities of the states under consideration do not
differ significantly between the EMA and the full eight-band model,
the electron and hole eigenenergies are modified in the order of 1 to 5~meV
between the two models (cf. table~\ref{tab:multibandEnergies}). The EMA is therefore able to
provide a good qualitative description of the charge densities of
electrons and holes. The eigenvalues do similarly agree well between EMA and
eight band $\mathbf{k}\cdot\mathbf{p}$ models.
% However, it can be seen
%that the inclusion of SO coupling induces an energetical difference between
%the first and the second excited hole state, and the SO coupling is essential
%for an accurate description of FSS, thus making the use of the more sophisticated
%eight band model mandatory.
%We have, nevertheless, used the computationally
%cheaper EMA to provide a rough estimate of the number of localised
%hole states inside the QD. We find of order 20 (doubly degenerate)
%confined hole states in the QD using the EMA. We note that the EMA
%tends to overestimate the splitting between different hole levels,
%so that the EMA therefore provides a lower bound on the number of
%states expected from a full calculation. The number
%of localised hole states in the systems under consideration
%is large compared to previous studies on
%site-controlled, (111)-oriented InGaAs QDs, where only 3 localised
%hole states were reported~\cite{KaDu10}. This difference arises due to
%the large base length of the QD in the present study.
%Given that we find only one (doubly degenerate) electron state
%confined in the QD under consideration, we therefore conclude that there will be
%a strong asymmetry between the number of confined electron and hole
%states in typical large-base site-controlled QDs. Furthermore,
Since the first four hole states are almost exclusively
$|X'\pm\rangle$ and $|Y'\pm\rangle$-like with negligible
$|Z'\pm\rangle$-like components ($\pm$ indicates spin-up and spin-down, respectively),
the here calculated bound hole states are predominantely of HH character.
This is in contrast to previous findings~\cite{KaDu10}, where the hole ground state is
89\% HH-like while the first excited state is a LH-like state (91\%).
We attribute this difference to the difference in the QD dimensions; the very
small aspect ratio here leads to the increased HH character for the highest
valence states.
%Therefore, we do not only find
%differences in the number of bound states but also in the character
%of the different states compared to the results found in
%Ref.~\cite{KaDu10} for site-controlled (111)-oriented InGaAs QDs.

% ---

Once the electronic structure of the site-controlled (111)-oriented
QDs is known, one can start to analyse the optical properties of
these systems. This task is beyond the scope of the present study.
However, from the electronic structure results derived here, one can
already start to discuss the procedure necessary to accurately
describe the optical properties, such as the FSS, of the system
under consideration.
%The fact that one is left with only one or two
%bound electron states but many closely-spaced hole states in
%realistic (111)-oriented site-controlled InGaAs QDs has far-reaching
%consequences for the theoretical analysis of the FSS and therefore
%the optical properties.
For an accurate description of the FSS,
configuration interaction (CI) calculations are often
applied~\cite{BeNa2003,SeSc2005,BaSc2005,Schl2007}. Such an approach
takes not only the direct Coulomb interaction between the carriers
into account, it accounts also for exchange and correlation
contributions. In the CI scheme, the many-body Hamiltonian is
expanded in the basis of anti-symmetrised products of bound
single-particle electron and hole states. In the case of
(001)-oriented InGaAs QDs, the CI scheme has been successfully
applied to study the FSS in these systems~\cite{SeSc2005,BeNa2003}.
However, the accuracy of this approach depends strongly on the
number of bound states taken into account in the
expansion~\cite{WiNa2006}. For example, Bester~\emph{et
al.}~\cite{BeNa2003} included 12 electron and 12 hole
single-particle states in the CI expansion to obtain reliable
results for the FSS in (001)-oriented InGaAs QDs.
Our calculations indicate all electronic states are close to
the GaAs band edges, such that only a small number of localised charge carriers
can be expected. It may thus become important to employ other techniques that
do not rely on a larger number of localised electronic states.
%Therefore, since
%one or two bound electron states are not sufficient for a reliable
%CI expansion, other techniques have to be applied in the case of
%realistic (111)-oriented site-controlled InGaAs QDs.

One strategy to circumvent the problems arising from a small number of bound
electron and hole states is to perform first a self-consistent
Hartree-Fock calculation and use the results from this calculation
as an input for the CI approach~\cite{KiKa2010}. Such calculations
could address issues such as the difference in binding energy of
excitons and biexcitons. However, it should be noted that disorder
effects can break the $C_{3v}$ symmetry in a (111)-oriented QD and
are therefore critical to determining the value of the FSS. We note
that there will be alloy-related disorder in the InGaAs QD layer. As
the energy differences between the hole states in the dot considered
here are only of the order of a few meV, it can be expected that
disorder effects may lead to a mixing of the different hole states
presented in figure~\ref{fig:el111}, even in a single-particle
picture~\cite{WaGo2011}. These disorder effects will therefore need to be included
in more detailed calculations of the dependence of FSS on dot size,
shape and composition. Thus, in addition to the very challenging
task to calculate the electronic structure of realistic
(111)-oriented, site-controlled InGaAs QDs, the accurate description
of the optical properties will be even more challenging, due to the
self-consistent cycles and disorder effects. The symmetry adapted
$\mathbf{k}\cdot\mathbf{p}$ formalism presented here then provides
one of the key building blocks to address this problem.

\section{Conclusion}
\label{sec:summary}

We have presented a (111)-rotated eight-band
$\mathbf{k}\cdot\mathbf{p}$ model for the description of ZB QDs
grown on a (111)-oriented surface. We have applied our model to the case of
a site-controlled In$_{0.25}$Ga$_{0.75}$As/GaAs QD with its
experimentally observed small aspect ratio to calculate electron and
hole single-particle states in these systems. Our approach yields a
significant reduction of the computational effort, by using a
Hamiltonian that is adapted to the specific properties of
(111)-oriented supercells. A detailed study showing the influence of
strain and polarisation potentials on the electronic properties has been performed.
%and yields a qualitative influence of these contributions on the electronic properties.
A recently published set of first- and second-order
piezoelectric constants does not significantly alter the outcome of
our calculations in comparison to the previous parameter set. A
comparison to the effective mass
%and six-band $\mathbf{k}\cdot\mathbf{p}$
simplification revealed the qualitative
ability of less sophisticated models to provide a good description
of electron and hole charge densities, while some
modifications in the eigenenergies occur for the case of the EMA,
outlining the importance of such an eight-band model.
%Nevertheless,
%we were able to use the EMA to identify the strong asymmetry in the
%number of confined electron and hole states in site-controlled,
%(111)-oriented QDs. We have discussed and outlined a formulation
%which takes this asymmetry into account and is suited to the
%analysis of optical properties, such as the FSS, in novel
%(111)-oriented III-V semiconductor nanostructures.
Our results highlight the need for a symmetry adapted approach as a first step
to calculate the electronic structure of such systems. We conclude
that our rotated eight-band $\mathbf{k}\cdot\mathbf{p}$ model is
well suited to a description of realistic, (111)-oriented ZB QDs and
can be used in combination with our previous work on the
correspondingly rotated continuum elasticity model, to carry out
broad studies on various possible modifications of realistic
site-controlled, (111)-oriented InGaAs/GaAs QDs, as well as
on other materials that exhibit a ZB crystal lattice.
Our work in combination with the $\mathbf{k}\cdot\mathbf{p}$ module of the
S/Phi/nX software library provides a ready-to-use approach to the electronic
properties of (111)-oriented ZB QDs, allowing in general for a reliable
and computationally inexpensive simulation of these novel nanostructures.

\ack
This work was carried out with the financial support of Science
Foundation Ireland (project numbers 06/IN.1/I90 and 10/IN.1/I2994).
We thank E. Pelucchi for useful discussions.

\appendix

\section{ Eight-band $\mathbf{k}\cdot\mathbf{p}$ Hamiltonian in the (001)-system}
\label{ap:kphamilconv}

The eight-band $\mathbf{k}\cdot\mathbf{p}$ Hamiltonian in a (001)-ZB
structure, expanded into basis states with symmetry
\mbox{$(|S\uparrow\rangle,|X\uparrow\rangle,|Y\uparrow\rangle,|Z\uparrow\rangle,|S\downarrow\rangle,|X\downarrow\rangle,|Y\downarrow\rangle,|Z\downarrow\rangle)$}
can be written in the following block matrix form~\cite{Schl2007}:
\begin{equation}
 H^{(001)}=
\begin{pmatrix}
 M(\mathbf{k}) & \Gamma\\
 -\Gamma^{*} & M^{*}(\mathbf{k})
\end{pmatrix}
\,\, ,
\end{equation}
where $M(\mathbf{k})$ and $\Gamma$ are $4\times$4 matrices. The
complex conjugate is denoted by $^{*}$. $M(\mathbf{k})$ can be
divided into four sub-matrices $M_\text{pe}$ (potential energy part;
terms independent of and linear in $\mathbf{k}$), $M_\text{ke}$
(kinetic energy part; terms quadratic in $\mathbf{k}$),
$M_\text{str}$ (strain dependent part) and $M_\text{so}$ (SO part):
%\begin{equation}
 $M(\mathbf{k})=M_\text{pe}+M_\text{ke}+M_\text{str}+M_\text{so}$ .
%\end{equation}
The matrix $M_\text{pe}$, describing terms independent of and linear
in $\mathbf{k}$, is given by:
\begin{eqnarray}
 M_\text{pe} &=&
\begin{pmatrix}
 E_\text{cb} & \text{i}Pk_x & \text{i}Pk_y & \text{i}Pk_z\\
-\text{i}Pk_x & \tilde{E}_\text{vb} & 0 & 0 \\
-\text{i}Pk_y & 0 & \tilde{E}_\text{vb} & 0 \\
-\text{i}Pk_z & 0 & 0 & \tilde{E}_\text{vb} \\
\end{pmatrix}\\
%&=&
%\left(\begin{array}{c|ccc}
% E_\text{cb} &  & {M^\text{cv}}^\dagger & \\\hline
% &  &  &  \\
%M^\text{cv} &  & M_{\tilde{E}_\text{vb}} & \\
% &  &  &  \\
%\end{array}\right)\,\, .
\end{eqnarray}
The matrix includes the CB edge energy $E_\text{cb}$ and the average
VB edge energy $\tilde{E}_\text{vb}$, described by the
$3\times3$ matrix $M_{\tilde{E}_\text{vb}}$. $E_\text{cb}$ and
$\tilde{E}_\text{vb}$ are defined in equation~(\ref{eq:Ec}). The Kane
coupling parameter $P$ is defined in equation~(\ref{eq:KaneP}).

The kinetic energy part of $M(\mathbf{k})$ for the  eight-band
Hamiltonian $H^{(001)}$ is given by $M_\text{ke}$:
\begin{eqnarray*}
M_\text{ke} &=&
\left(\begin{array}{c|ccc}
A'\mathbf{k}^2 & 0 & 0 & 0\\\hline
0 &  &  & \\
0 &  & H^\text{VB} & \\
0 &  &  &
\end{array}\right)
\end{eqnarray*}
The parameter $A'$ is defined in equation~(\ref{eq:KaneA}). Following
Schliwa~\emph{et al.}~\cite{ScWi2007}, we neglect the quadratic
coupling between the CB and VBs which is related
to the Kane parameter $B$.

The VB part is described by the $3\times3$ matrix $H^\text{VB}$:
%describes the valence band without SO-coupling and strain and in the $(|X\rangle,|Y\rangle,|Z\rangle)^T$ basis is given by
\begin{align*}
&H^\text{VB}=\\
&
\left(\begin{array}{ccc}
\left\{\begin{array}{c}
 \tilde{l}k^2_x+\\
m\left(k^2_y+k^2_z\right)
\end{array}\right\}
  & \tilde{n}k_xk_y & \tilde{n}k_xk_z\\
 \tilde{n}k_xk_y &
\left\{\begin{array}{c}
\tilde{l}k^2_y+\\
m\left(k^2_x+k^2_z\right)
\end{array}\right\}
& \tilde{n}k_yk_z\\
 \tilde{n}k_xk_z & \tilde{n}k_yk_z &
\left(\begin{array}{c}
\tilde{l}k^2_z+\\
m\left(k^2_x+k^2_y\right)
\end{array}\right)
\end{array}\right)
\end{align*}
with $\tilde{l}$, $m$ and $\tilde{n}$ given by
\begin{equation*}
\tilde{l} = \frac{P^2}{E_g}-\frac{1}{2}(\gamma_1+4\gamma_2)\, , \,  m = -\frac{1}{2}(\gamma_1-2\gamma_2)\, , \, \tilde{n} = \frac{P^2}{E_g}-3\gamma_3\, ,
\end{equation*}
where $E_g$ is the fundamental band gap of the material under
consideration and $\gamma_i$ are the Luttinger parameters, given in
units of $\hbar^2/m_0$, with $m_0$ being the mass of the electron.

The strain dependent part $M_\text{str}$ is given by:
\begin{align*}
& M_\text{str} =\\
&\begin{pmatrix}
 a_cTr(\epsilon) & -\text{i}P\epsilon_{1\beta}k^\beta & -\text{i}P\epsilon_{2\beta}k^\beta & -\text{i}P\epsilon_{3\beta}k^\beta\\
\text{i}P\epsilon_{1\beta}k^\beta &
\left(\begin{array}{c}
a_vTr(\epsilon)\\
+b\epsilon_{B,x}
\end{array}\right)
& \sqrt{3}d\epsilon_{12} & \sqrt{3}d\epsilon_{13} \\
\text{i}P\epsilon_{2\beta}k^\beta & \sqrt{3}d\epsilon_{12} &
\left(\begin{array}{c}
a_vTr(\epsilon)\\
+b\epsilon_{B,y}
\end{array}\right)
& \sqrt{3}d\epsilon_{23} \\
\text{i}P\epsilon_{3\beta}k^\beta & \sqrt{3}d\epsilon_{13} & \sqrt{3}d\epsilon_{23} &
\left(\begin{array}{c}
a_vTr(\epsilon)\\
+b\epsilon_{B,z}
\end{array}\right)\\
\end{pmatrix}\\
\end{align*}
with $\epsilon_{B,x}=2\epsilon_{11}-(\epsilon_{22}+\epsilon_{33})$,
$\epsilon_{B,y}=2\epsilon_{22}-(\epsilon_{11}+\epsilon_{33})$ and
$\epsilon_{B,z}=2\epsilon_{33}-(\epsilon_{11}+\epsilon_{22})$, and
where $\epsilon_{ij}$ denotes the different strain tensor
components. The hydrostatic deformation potential of the VB is
denoted by $a_v$ while $a_c$ denotes the hydrostatic deformation
potential of the CB. The uniaxial VB deformation potentials are
given by $b$ and $d$. Following Schliwa \emph{et
al.}~\cite{ScWi2007}, we neglect shear-strain related CB-VB
coupling. The SO coupling is described by the matrices $M_\text{so}$
and $\Gamma_\text{so}$, which are identical to $M'_\text{so}$, and
$\Gamma'_\text{so}$, given in equation~(\ref{eq:SpinDiag}).

%!!!!!!!!!!!!!!!!!!!!!!!!!!!!!!!!!!!!!!!!!!!
% \bibliography{../phdstef}

\bibliographystyle{unsrt.bst}

\end{document}